\documentclass[twocolumn,english,aps,pre,groupedaddress,showpacs,showkeys]{revtex4-1}
\usepackage[latin9]{inputenc}
\usepackage{color}
\usepackage{bm}
\usepackage{amsmath}
\usepackage{amssymb}
\usepackage{graphicx}
\usepackage{esint}

\makeatletter

\usepackage{babel}
\newcommand{\ve}[1]{\ensuremath{\mbox{\boldmath$#1$}}}
\newcommand{\mb}[1]{{\mathbf{#1}}}
\newcommand{\A}{\ve{\tilde{\Gamma}}}

\usepackage{babel}

\usepackage{babel}

\makeatother

\usepackage{babel}

\usepackage{hyperref}
\usepackage{cleveref}
\begin{document}

\title{Is the kinetic equation for turbulent gas-particle flows ill-posed? }

\author{M.~Reeks$^{\#}$}

\email[]{mike.reeks@ncl.ac.uk}

\author{D.C.~Swailes$^{\#\#}$}

\affiliation{School of Mechanical \& Systems Engineering $^{\#}$and School of
Maths, Statistics \& Physics$^{\#\#}$, Newcastle University, Newcastle
upon Tyne NE1 7RU, UK}

\author{A.~Bragg}

\email{andrew.bragg@duke.edu}

\affiliation{Department of Civil \& Environmental Engineering, Duke University,
USA}

\date{\today}
\begin{abstract}
This paper is about well-posedness and realizability of the kinetic
equation for gas-particle flows and its relationship to the Generalized
Langevin Model (GLM) PDF equation. Previous analyses claim that this
kinetic equation is ill-posed, that in particular it has the properties
of a backward heat equation and as a consequence, its solutions will
in the course of time exhibit finite-time singularities. We show that
the analysis leading to this conclusion is fundamentally incorrect
because it ignores the coupling between the phase space variables
in the kinetic equation and the time and particle inertia dependence
of the phase space diffusion tensor. This contributes an extra $+ve$
diffusion that always outweighs the contribution from the$-ve$ diffusion
associated with the dispersion along one of the principal axes of
the phase space diffusion tensor. This is confirmed by a numerical
evaluation of analytic solutions of these $+ve$ and $-ve$ contributions
to the particle diffusion coefficient along this principal axis. We
also examine other erroneous claims and assumptions made in previous
studies that demonstrate the apparent superiority of the GLM PDF approach
over the kinetic approach. In so doing we have drawn attention to
the limitations of the GLM approach which these studies have ignored
or not properly considered, to give a more balanced appraisal of the
benefits of both PDF approaches. 
\end{abstract}
\maketitle

\section{INTRODUCTION\label{sec:INTRODUCTION}}

The Probability Density Function (PDF) approach has proved very useful
in studying the behavior of stochastic systems. Familiar examples
of its usage occur in the study of Brownian Motion \cite{Chandrasekhar43}
and in the kinetic theory of gases \cite{Chapman&Cowling}. In more
recent times it has been used extensively by Pope and others to model
turbulence \cite{Pope86} and turbulence related phenomena such as
combustion \cite{Pope91} and atmospheric dispersion \cite{MacInnes&Bracco92}.
This paper is about its application to particle transport in turbulent
gas-flows where it has been been developed and refined over a number
of years by numerous authors. It has been successfully applied to
a whole range of turbulent dispersed flow problems involving mixing
and dispersion as well as particle collisions and clustering in a
particle pair formulation of the approach. It has also formed a fundamental
basis for dealing with complex flows in formulating the continuum
equations and constitutive relations for the dispersed phase precisely
analogous to the way the Maxwell Boltzmann equation has been used
in the kinetic theory. It has become an established technique for
studying dispersed flows so much so that the method and its numerous
applications are the subject of a recent book \cite{ZaichikAliphenkovSinaiskibook}
and the subject of a chapter in the recent  Multiphase flow Handbook
\cite{ReeksSimoninFede}

There are currently two PDF approaches that have been used extensively
to describe the transport, mixing and collisions of small particles
in turbulent gas flows. The first approach referred to as the kinetic
approach, is based on a kinetic equation for the PDF $p(\ve x,\ve v,t)$
of the particle position $\ve x$ and velocity $\ve v$ at time $t$.
This equation based on a particle equation of motion involving the
flow velocity along a particle trajectory derived from a Gaussian
stochastic flow field. In the kinetic equation the particles' random
motion arisng from this stochastic field is manifest as a diffusive
flux which is a linear combination of gradient diffusion in both $\ve x$
and $\ve v$. Transient spatio-temporal structures in the turbulence
give rise to an extra force due to clustering and preferential sweeping
of particles \cite{Maxey1987}.

In the second PDF approach an equation for the PDF $p(\ve x,\ve v,\ve u,t)$
is constructed, where $\ve u$ is the carrier flow velocity sampled
along particle paths. Thus, unlike the kinetic approach, the flow
velocity in this approach is retained in the particle phase space,
and is described by a model evolution equation. In particular, this
PDF model is based on a generalized Langevin model (GLM) (see Pope
\cite{Pope86}) where the velocity of the underlying carrier flow
measured along a particle trajectory is described by a generalized
Langevin equation. As such the associated PDF equation is described
by a Fokker-Planck equation. This GLM PDF equation has sometimes been
inappropriately referred to as the dynamic PDF equation \cite{minier15}
implying that it is a more general PDF approach from which the kinetic
equation can in general be derived. However it is important to appreciate
the kinetic equation is not a standard Fokker-Planck equation, since
it captures the non-Markovian features of the underlying flow velocities.
\\
 \\
The problem of closure and the associated realizability and well posedness
of PDF equations are profoundly important in the study of stochastic
equations. So despite the successful application of the kinetic equation
to a whole range of problems, recent claims in the literature of ill-posedness
and realizability of this equation are disturbing and a serious concern.
The root cause of this concern is the non-positive definiteness of
the diffusion tensor associated with the phase space diffusion flux.
That in particular this tensor has both $+ve$ and $-ve$ eigenvalues
implying that along the eigenvectors with a $-ve$ eigenvalue the
particle dispersion exhibits the properties of a backward diffusion
equation leading to solutions with finite time singularities. In fact
Minier and Profeta \cite{minier15} following a detailed analysis
of the relative merits of the 2 PDF approaches, have concluded that
the kinetic equation is ill-posed and therefore an invalid description
of disperse two-phase flows (except in the limiting case for particles
with large Stokes numbers when the kinetic equation reduces to a Fokker-Planck
equation). This raises a number of issues and inconsistencies that
we wish to examine and resolve:
\begin{enumerate}
\item The closure of the diffusive terms in the kinetic equation is exact
for a Gaussian process for the aerodynamic driving forces in the particle
equation of motion. Not withstanding any $-ve$ eigenvalues, such
dispersion processes are demonstrably forward rather than backward
in time with statistical moments that monotonically increase rather
than decrease with time. This behaviour is reflected in the analytic
solutions of the kinetic equation for particle dispersion in shear
flows in which the mean shear is linear and the turbulence is statistically
homogeneous and stationary (see Hyland et al \cite{hyland99}, Swailes
and Darbyshire \cite{Darbyshire_Swailes96}). In these generic flows,
there is exact correspondence of the analytical solution with a random
walk simulation using a Lagrangian particle tracking approach, solving
the individual particle equations of motion in the associated Gaussian
random flow field. See as an example the illustration in Figure \ref{fig:simple turbulent shear flow }.
\begin{figure}
\noindent \begin{centering}
\includegraphics[bb=0bp 200bp 595bp 700bp,scale=0.3]{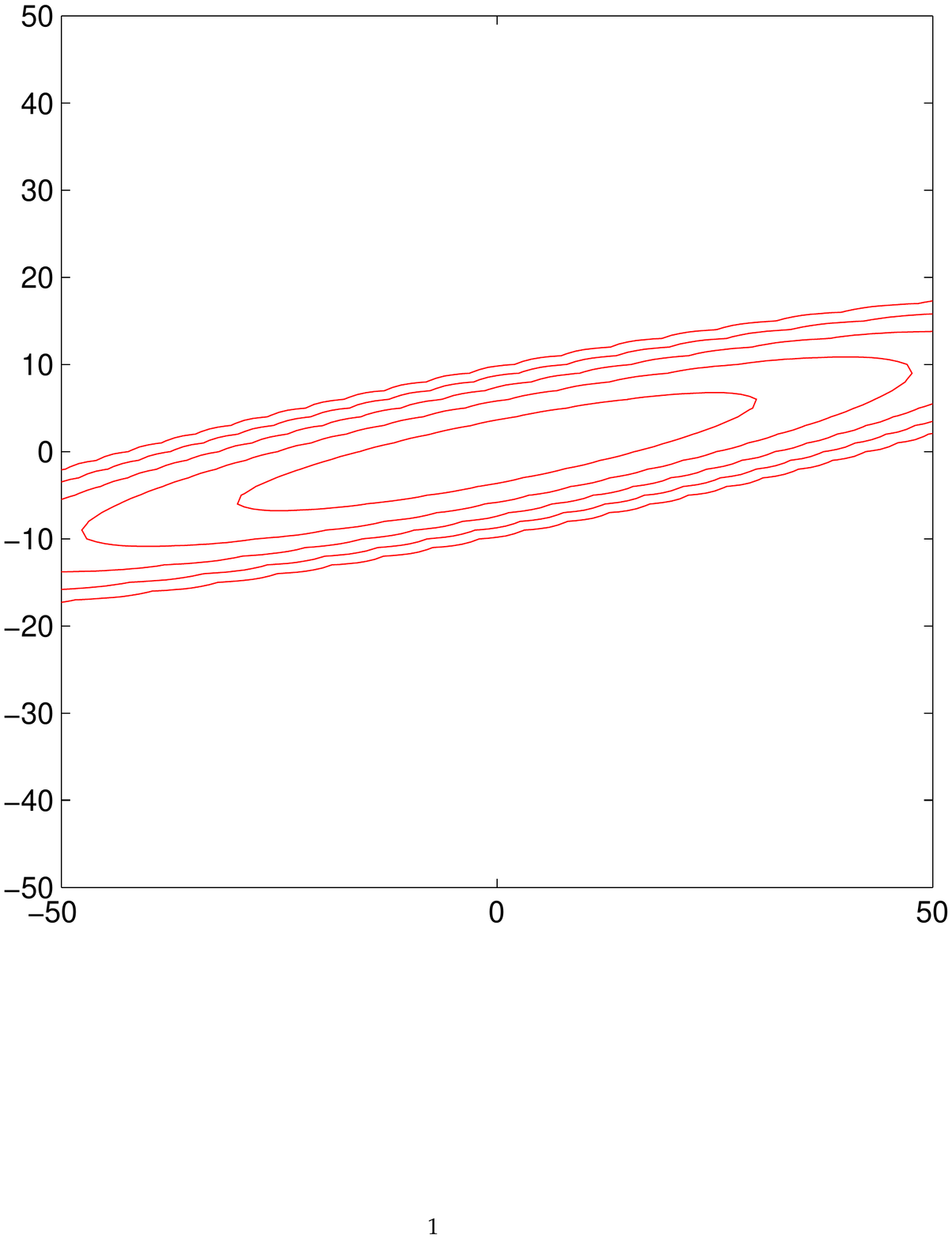}
\par\end{centering}

\noindent \begin{centering}
Analytic solution 
\par\end{centering}

\noindent \begin{centering}
\includegraphics[bb=0bp 200bp 595bp 700bp,scale=0.3]{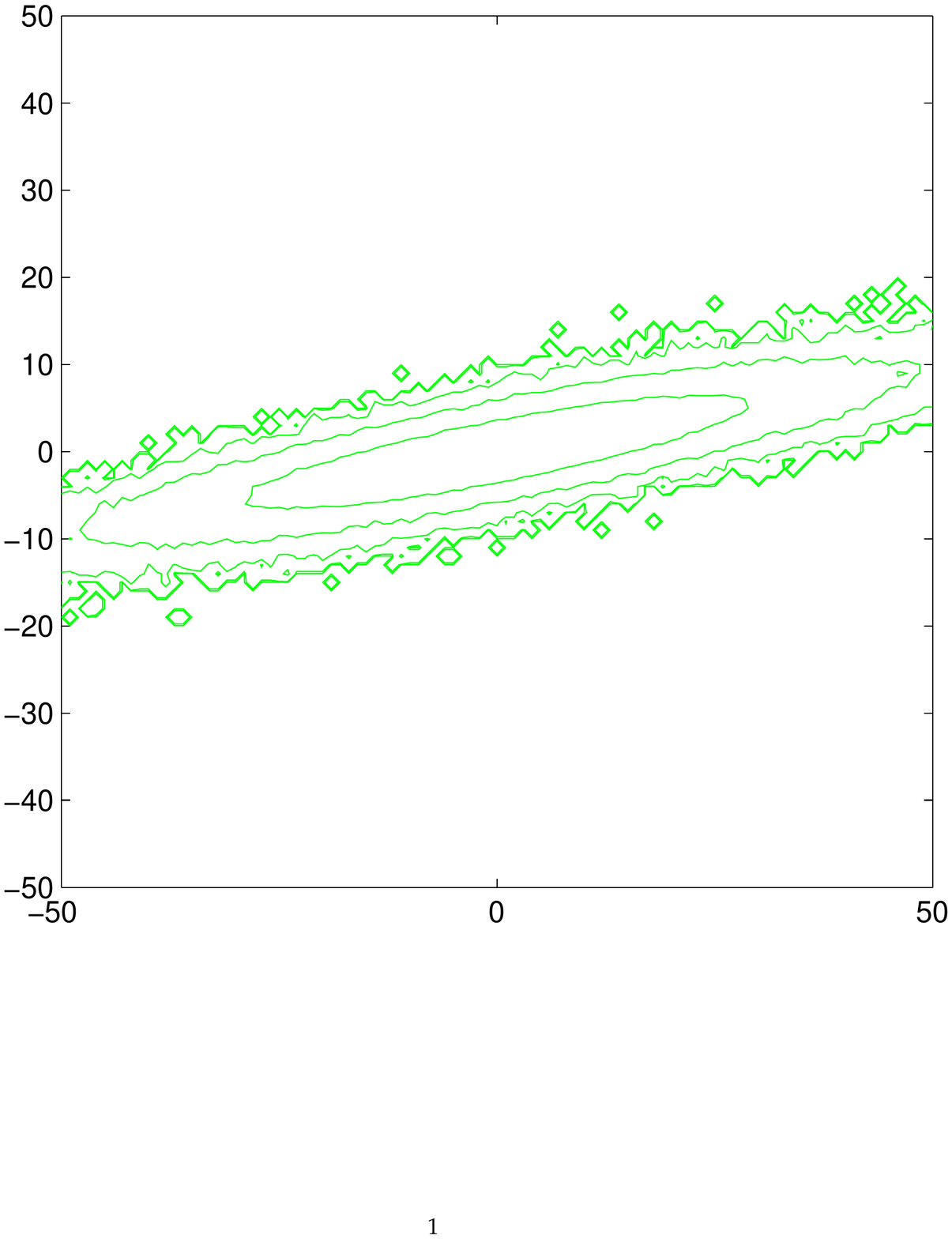}
\par\end{centering}

\noindent \begin{centering}
Random walk simulation\\

\par\end{centering}

\begin{centering}
\includegraphics[scale=0.4]{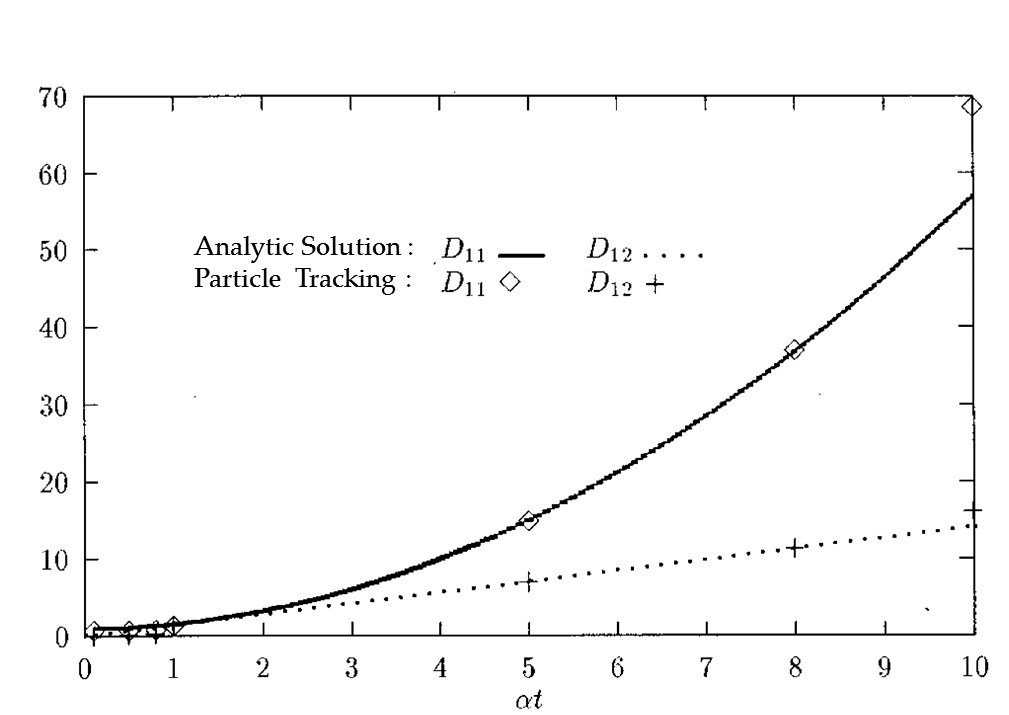}
\par\end{centering}

\begin{centering}
Diffusion coefficients $D_{ij}$ as a function of time $t$, $\alpha$=
shear rate
\par\end{centering}

\noindent \centering{}\protect\caption{\label{fig:simple turbulent shear flow } Dispersion of an instantaneous
point source of particles in a simple shear flow. Comparison of the
analytic solution of the kinetic equation for the particle spatial
concentration and a random walk simulation based on Stokes drag with
a Gaussian process for the aerodynamic driving force. For more precise
details see \cite{Darbyshire_Swailes96,hyland99,ReeksSimoninFede} }
\end{figure}

\item In simple generic flows the GLM PDF equation is entirely consistent
with the kinetic equation i.e the kinetic equation is recoverable
from the GLM equations and has exactly the same solution for the same
mean flows and statistical correlations for the turbulent velocity
$\ve u$ along particle trajectories. They are both compatible with
a Gaussian process. The claim of ill posedness of the kinetic equation
would therefore seem to contradict the well posedness associated with
the Fokker-Planck equation of the GLM.
\end{enumerate}
So the first objective of the analysis we present here is to show
that despite the non positive definiteness of the phase space diffusion
tensor, this does not imply backward diffusion and the existence of
finite time singularities, that the kinetic equation is well posed
and has realizable solutions that are forward rather than backward
in time consistent with a Gaussian process. We shall show that this
is intimately related to the non Markovian nature of the kinetic equation,
that the time evolution of the phase dispersion tensor from its initial
state and the coupling between phase space variables are crucial considerations.
In the course of this analysis we will recall the stages of the development
of the kinetic equation and the important role played by certain consistency
and invariance principles which taken together with the other features
determining well-posedness and realizability  have not been properly
understood or appreciated in previous analyses.\\
\\
Previous work has purported to show that the GLM is a more general
approach than the kinetic approach. That in particular the kinetic
equation can be derived from the GLM, and that the features of transport
and mixing in more general non uniform inhomogeneous turbulent flows
implicit in the solutions of the kinetic equation are intrinsic to
the GLM. So the second objective of this analysis is to examine the
basis for this assertion. In the process, we provide a more balanced
appraisal of the benefits of both PDF approaches\textcolor{black}{{}
and point out the limitations of the GLM that have been ignored in
previous analyses. We regard these limitations to be areas for improvement
of the GLM rather than inherent deficiencies. }Like all modeling approaches,
each of the two approaches considered have their strengths and weaknesses.
A categorical dismissal of one in preference to another in previous
work would seem misplaced. From a practical point of view this paper
is more about how one approach can support the other in solving dispersed
flow problems.

\section{Ill-Posed Kinetic PDF Equations?}

In this section we examine in detail the previous analysis of Minier
\& Profeta (M\&P) \cite{minier15} that leads to the assertion of
ill-posedness of the kinetic equation. For ease of comparison we use
the same notation here and throughout the paper. Thus M\&P consider
particle phase-space trajectories $\ve Z_{p}(t)=(\ve X_{p}(t),\ve U_{p}(t))$
governed by 
\begin{equation}
\dot{\ve X}_{p}=\ve U_{p},\hspace{2em}\dot{\ve U}_{p}=\frac{1}{\tau_{p}}(\ve U_{s}-\ve U_{p})+\ve F_{ext}.\label{pem1}
\end{equation}
$\ve U_{s}(t)$ representing a flow velocity at time $t$ sampled
along the trajectory $\ve X_{p}(t)$, and $\ve F_{ext}$ an external
body force e.g. gravity. In the kinetic modeling framework, $\ve U_{s}$
is derived via an underlying flow velocity field $\ve u_{f}(\ve x,t)$
which has both a mean $\langle\ve u_{f}\rangle$ and fluctuating (zero
mean component) $\ve u_{f}^{\prime}$. That is $\ve U_{s}=\ve u_{f}(\ve X_{p}(t),t)$.
Treating $\ve u_{f}(\ve x,t)$ as a Gaussian stochastic flow field,
and with the particle response time $\tau_{p}$ as a constant independent
of the particle Reynolds no (i.e Stokes relaxation), the PDF $p(\ve z,t)$
defining the distribution of $\ve Z_{p}(t)$ then satisfies a transport
equation (the kinetic equation) which can be written compactly in
phase-space notation as\_ 
\begin{equation}
\partial_{t}p=-\partial_{\ve z}\cdot\ve ap\ +\ \tfrac{1}{2}\partial_{\ve z}\cdot\left(\partial_{\ve z}\cdot\ve Bp\right)\label{pde1}
\end{equation}
where $\ve z=(\ve x,\ve v)$ refers to the particle position and velocity
(in a fixed frame of reference) and 
\begin{equation}
\ve a=\left(\ve v,\ve F_{ext}+\dfrac{1}{\tau_{p}}\bigl(\langle\ve u_{f}(\ve x,t)\rangle-\ve v\bigr)+\bm{\kappa}\right)\label{a}
\end{equation}
\begin{equation}
\ve B=\left(\begin{array}{c|c}
\bm{0} & \bm{\lambda}\\
\hline \bm{\lambda}^{\top} & \bm{\mu}+\bm{\mu}^{\top}
\end{array}\right)\label{B}
\end{equation}
$\ve\lambda$ and $\ve\mu$ are diffusion tensors that define gradient
dispersion separately in real space ($\ve x)$ and velocity space
$(\ve v)$ respectively. They are functions of time and depend on
the particle response to the carrier flow velocity fluctuations along
its trajectory. The specific forms for $\ve\lambda$ and $\ve\mu$
based on the LHDI closure scheme \cite{Reeks92} 
\begin{align*}
\ve\lambda & =\tau_{p}^{-2}\int_{0}^{t}\ve g^{T}(t-s)\cdot\langle\ve u_{f}^{\prime}(\ve x,\ve v,t\mid s)\ve u_{f}{}^{\prime}(\ve x,t)\rangle ds\\
\ve\mu & =\tau_{p}^{-2}\int_{0}^{t}\dot{\ve g^{T}}(t-s)\cdot\langle\ve u_{f}^{\prime}(\ve x\mathbf{,}\ve v,t\mathbf{\mid}s)\ve u_{f}{}^{\prime}(\ve x,t)\rangle ds\\
\ve\kappa & =-\tau_{p}^{-2}\int_{0}^{t}\ve g^{T}(t-s)\cdot\langle\ve u_{f}^{\prime}(\ve x\mathbf{,}\ve v,t\mathbf{\mid}s)\partial_{\ve x}\ve u_{f}^{\prime}(\ve x,t)\rangle ds
\end{align*}
where the particle response tensor $\ve g(t-s)$ has elements $g_{ij}(t\mid s)$
corresponding to the displacement at time $t$ in the $j$-direction
when $\mbox{\ensuremath{\tau}}_{p}u_{f}^{\prime}$ is an impulsive
force $\delta(t-s)$ applied in the $i$-direction. In general $\ve g(t-s)$
depends up the local straining and rotation of the flow. Note the
response tensor based on the Furutsu-Novikov closure scheme \cite{swailes97}
is slightly different in definition (see Bragg \& Swailes \cite{SwailesBragg2012}
for a discussion of the  different closure schemes for the kinetic
equation). Following the analysis of M\&P, we consider the case for
dispersion of an instantaneous point source in statistically stationary
homogeneous and isotropic turbulence with a zero external force $\ve F_{ext}=0$
, in which case $\ve g(t)=(1-e^{-t/\tau_{p}})\:\ve I$ and 
\begin{align*}
\ve\lambda & =\tau_{p}^{-2}\int(1-e^{-s/\tau_{p}})R(s)ds\:\ve I\\
\ve\mu & =\tau_{p}^{-2}\int e^{-s/\tau_{p}}R(s)ds\:\ve I\\
\ve\kappa & =0
\end{align*}
where $R(s)$ is the autocorrelation $\frac{1}{3}\left\langle \ve U_{s}^{\prime}(0)\cdot\ve U_{s}^{\prime}(s)\right\rangle $
of the flow velocity fluctuations $\ve U^{\prime}(s)$ measured along
a particle trajectory. \eqref{pde1},\eqref{a}), \eqref{B} correspond
to equations (65), (66), (67) in \cite{minier15}. M\&P claim that
equation \eqref{pde1} is ill-posed in the sense that solutions to
this can (will) exhibit unphysical behaviour except in special or,
to use their phrase, `lucky' cases. Specifically, they assert that
solutions $p$ of equation \eqref{pde1} will exhibit finite-time
singularities except for very special initial conditions, for example
with a Gaussian form. Their justification for this claim is based
on an analysis centered round the observation that $\ve B$ is not
positive-definite but possesses both negative and positive eigenvalues.
We show here that their analysis is incorrect.

Firstly we note that equation \eqref{pde1} is \emph{not} a model
for the PDF of $\ve Z_{p}(t)$, but describes precisely how this PDF
must evolve. There is an \emph{exact} correspondence between equation
\eqref{pde1} and the underlying equation of motion \eqref{pem1}.
This equivalence, \textit{i.e.}~ the formal derivation of \eqref{pde1}
from \eqref{pem1}, is subject only to the requirement that the field
$\ve u_{f}(\ve x,t)$ is Gaussian. Then, notwithstanding the non-definiteness
of $\ve B$, equation \eqref{pde1} is an exact description of how
$p$, as determined by \eqref{pem1}, behaves. Contrary to previous
claims \cite{minier15} no Gaussian (or other) constraint is necessary
on the initial distribution $p^{0}(\ve z)$ of $\ve Z_{p}(0)$. Thus,
should solutions to \eqref{pde1} exhibit finite-time, or even asymptotic
($t\rightarrow\infty$), singularities when $p^{0}$ is non-Gaussian,
then this feature must be inherent in the system determined by \eqref{pem1}.
Either this singular behaviour is intrinsic to the system, or the
analysis upon which M\&P base their conclusion is incorrect.

To demonstrate that the non-definiteness of $\ve B$, coupled with
arbitrary initial conditions, does not lead to singular solutions
of equation \eqref{pde1} we note that the solution to this equation
can be written 
\begin{equation}
p(t;\ve z)=\int\phi(t;\ve z,\ve z^{\prime})p^{0}(\ve z^{\prime})d\ve z^{\prime}\label{superposition}
\end{equation}
where $\phi(t;\ve z,\ve z^{\prime})$ is the fundamental solution
satisfying $\phi(0;\ve z,\ve z^{\prime})=\delta(\ve z-\ve z^{\prime})$.
Now consider the case when $\ve{U}_{s}^{\prime}(t)=\ve{u_{f}}^{\prime}(\ve{X}_{p},t)=\ve{u_{f}}(\ve{X}_{p},t)-\langle\ve{u_{f}}\rangle(\ve{X}_{p},t)$
is treated, \textit{ab initio}, as a Gaussian process. The structure
of equation \eqref{pde1} remains unchanged, except $\bm{\kappa}\equiv\bm{0}$
and $\bm{\lambda}$, $\bm{\mu}$ are independent of $\ve{Z}$ (but,
crucially, they will still depend on $t$). $\ve{B}$ still has negative
eigenvalues. With $\langle\ve u_{f}\rangle$ linear in $\ve x$ (and
$\ve{F}_{\mathrm{ext}}$ constant) the form of $\phi$ is well-documented,
both in general terms and for a number of specific linear flows \cite{reeks05,Darbyshire_Swailes96,hyland99}.
This solution is Gaussian, and it is straightforward to show that
it corresponds exactly, as it must, to the Gaussian form of $\ve{Z}_{p}$
determined by equation \eqref{pem1}. Thus, any singular behaviour
of the general solution $p$, defined by equation \eqref{superposition},
can only be a consequence of degeneracy in the Gaussian form of $\phi$,
and not the form of an arbitrary initial distribution $p^{0}$. Again,
should such degeneracy exist then it would be symptomatic of behaviour
determined by \eqref{pem1}, and not some artifact of the non-definiteness
of $\ve{B}$.\\

There are several flaws in the analysis upon which M\&P base their
claim of ill-posedness: To begin, they consider a form of equation
\eqref{pde1} in which $\ve{B}$ is taken as independent of time,
arguing that this corresponds to stationary isotropic turbulence.
This is not correct. $\ve{B}$ is intrinsically time dependent. This
dependence reflects the non-zero time correlations implicit in the
turbulent velocity field $\ve{U}_{f}$, and the consequent non-Markovian
nature of $\ve{Z}_{p}$. Moreover, and crucially, $\ve{B}(0)=\bm{0}$
unless the initial values $\ve{U}_{p}(0)$, $\ve{U}_{s}(0)$ are correlated.
A detailed analysis of this is given in \cite{hyland99}. So, even
when $\ve{B}\rightarrow\ve{B}^{\infty}$ (constant) as $t\rightarrow\infty$,
it is inappropriate to set $\ve{B}=\ve{B}^{\infty}$ in a formal analysis
of the time problem. Indeed, it is straightforward to show that the
fundamental solution $\phi$ breaks down for arbitrarily small $t$
when this inappropriate approximation is introduced.

Of course, the non-definiteness of $\ve{B}$ is not altered by taking
this tensor to be $t$ dependent. The eigensolution based transformation
that M\&P introduce can still be invoked. Analogous to equation (71)
in \cite{minier15} we define trajectories $\widetilde{\ve{Z}}_{p}(t)$
with components $(\widetilde{Z}_{p1},\widetilde{Z}_{p2}$) in a transformed
phase space $\widetilde{\ve z}=\left(\widetilde{z}_{1},\widetilde{z}_{2}\right)$
with 
\begin{equation}
\widetilde{\ve{Z}}_{p}(t)=\mathrm{P}^{\top}\cdot\ve{Z}_{p}(t)\label{eq:transform}
\end{equation}
where $\mathrm{P}(t)$ is the transformation matrix determined by
the (now time dependent) normalized eigenvectors of $\ve{B}$. Thus,
$\mathrm{P}^{\top}\cdot\mathrm{P}=\mathrm{I}$ and $\mathrm{P}^{\top}\cdot\ve{B}\cdot\mathrm{P}=\bm{\Lambda}=\mathrm{diag}(\omega_{i})$,
with $\omega_{i}$ the eigenvalues of $\ve{B}$. We note that, in
applying this to the 2D case considered by the M\&P, it is sensible
to label the two eigenvalues such that $\omega_{1}<0$, $\omega_{2}>0$
since this gives $\mathrm{P}(0)=\mathrm{I}$. By neglecting the time
dependence in $\ve{B}$ M\&P missed this point and chose the opposite
ordering (see equation (69) in \cite{minier15}). Here we take $\omega_{1}<0$.

In using the transform given by equation \eqref{eq:transform} it
is important to note that equation \eqref{pem1} governing $\ve{Z}_{p}(t)$
is not to be interpreted as a stochastic differential equation driven
by a white-noise process, and equation \eqref{pde1} is not a corresponding
Fokker-Planck equation. Clearly this would be nonsense since $\ve{B}$
is not positive-definite. It is more transparent (and correct) to
note that equation \eqref{eq:transform} implies that the PDF $\widetilde{p}(\widetilde{\ve{z}},t)$
of $\widetilde{\ve{Z}}_{p}(t)$ is related to the PDF $p(\ve{z},t)$
of $\ve{Z}_{p}(t)$ by $\widetilde{p}\,\vert\mathrm{J}\vert=p$, where
$\mathrm{J}=\det\bigl[\mathrm{P}\bigr]$ is the Jacobean of the transform
$\widetilde{\ve{z}}=\mathrm{P}^{\top}\cdot\ve{z}$. Since $\mathrm{P}$
is orthogonal we have $\mathrm{J}=1$. The PDF equation for $\widetilde{p}$
is 
\begin{equation}
\partial_{t}\widetilde{p}=-\partial_{\widetilde{\ve{z}}}\cdot\widehat{\ve{a}}\widetilde{p}\ +\ \tfrac{1}{2}\partial_{\widetilde{\ve{z}}}\cdot\left(\partial_{\widetilde{\ve{z}}}\cdot\bm{\Lambda}\widetilde{p}\right)\label{pde2}
\end{equation}
where $\widehat{\ve{a}}=\mathrm{P}^{\top}\cdot\widetilde{\ve{a}}+\mathrm{R}\cdot\widetilde{\ve{z}}$,
$\widetilde{\ve{a}}(\widetilde{\ve{z}},t)=\ve{a}(\ve{z},t)$, $\mathrm{R}=\dot{\mathrm{P}}^{\top}\cdot\mathrm{P}$.
This is analogous to equation (72) in \cite{minier15}, except these
authors have not included the time dependence in $\ve{B}$ and so
set $\dot{\mathrm{P}}=0$. We note that $\mathrm{R}$ represents a
rate of rotation matrix, $\mathrm{trace}(\mathrm{R})=0$. In the 2D
model considered, the authors integrate equation \eqref{pde2} over
$\widetilde{z}_{2}$ (corresponding to the transformed variable with
the positive eigenvalue $\omega_{2}$) to obtain (compare with equation
(74) in \cite{minier15}) 
\begin{equation}
\partial_{t}\widetilde{p}_{r}=-\partial_{\widetilde{\mathrm{z}}_{1}}\overline{\widehat{\mathrm{a}}}_{1}\widetilde{p}_{r}\ -\ \partial_{\widetilde{\mathrm{z}}_{1}}^{2}\tfrac{1}{2}\vert{\omega_{1}}\vert\widetilde{p}_{r}\label{pde3}
\end{equation}
where $\widetilde{p}_{r}$ is the PDF for $\widetilde{Z}_{p1}$ and
$\overline{\widehat{\mathrm{a}}}_{1}\widetilde{p}_{r}=\int\widehat{\mathrm{a}}_{1}\widetilde{p}\,d\widetilde{\mathrm{z}}_{2}$.
Based on the negative diffusion coefficient in equation \eqref{pde3}
M\&P seek to show that this equation and so also equation \eqref{pde1}
is ill-posed. Their argument fails to take into account that the conditional
average $\overline{\widehat{\mathrm{a}}}_{1}$ is a density weighted
average, i.e its value at $z_{1}$ is dependent upon the distribution
of $Z_{p2}(t)$ at $z_{1}$ which itself can be a function $z_{1}$.
For instance using a more explicit notation we may write 
\begin{equation}
\overline{\widehat{\mathrm{a}}}_{1}\equiv\Big\langle\widehat{\mathrm{a}}_{1}(\widetilde{\mathrm{z}}_{1},\widetilde{\mathrm{Z}}_{p2}(t))\Big\rangle_{\widetilde{z}_{1}}\label{eq:abardef}
\end{equation}
where $\langle\cdot\rangle_{\widetilde{z}_{1}}$ denotes an ensemble
average conditioned on $\widetilde{\mathrm{Z}}_{p1}(t)=\widetilde{\mathrm{z}}_{1}$.
What equation (\ref{eq:abardef}) illustrates is that only a sub-set
of all trajectories $\widetilde{\mathrm{Z}}_{p2}(t)$ contribute to
$\overline{\widehat{\mathrm{a}}}_{1}$, namely those that are also
associated with $\widetilde{\mathrm{Z}}_{p1}(t)=\widetilde{\mathrm{z}}_{1}$.
The term $\overline{\widehat{\mathrm{a}}}_{1}$ is therefore affected
by coupling between $\widetilde{\mathrm{Z}}_{p1}(t)$ and $\widetilde{\mathrm{Z}}_{p2}(t)$.
Indeed, in the case where $\widetilde{\mathrm{Z}}_{p1}(t)$ and $\widetilde{\mathrm{Z}}_{p2}(t)$
are statistically decoupled, we have 
\begin{equation}
\Big\langle\widehat{\mathrm{a}}_{1}(\widetilde{\mathrm{z}}_{1},\widetilde{\mathrm{Z}}_{p2}(t))\Big\rangle_{\widetilde{z}_{1}}=\Big\langle\widehat{\mathrm{a}}_{1}(\widetilde{\mathrm{z}}_{1},\widetilde{\mathrm{Z}}_{p2}(t))\Big\rangle,\label{abardef2}
\end{equation}
i.e. \emph{all} realizations of $\widetilde{\mathrm{Z}}_{p2}(t)$
would contribute to $\overline{\widehat{\mathrm{a}}}_{1}$. In this
case $\overline{\widehat{\mathrm{a}}}_{1}(z_{1})$ is convective as
M\&P have assumed. However, in general, $\widetilde{\mathrm{Z}}_{p1}(t)$
and $\widetilde{\mathrm{Z}}_{p2}(t)$ will be statistically coupled,
and as a consequence $\overline{\widehat{\mathrm{a}}}_{1}$ cannot
be treated as an arbitrary convective term. Indeed as we shall show
momentarily, the term $\overline{\widehat{\mathrm{a}}}_{1}$ is associated
with both convective and diffusive fluxes, and its diffusional contribution
offsets that associated with the negative eigenvalue.

By failing to appreciate this particular property of $\overline{\widehat{\mathrm{a}}}_{1}$,
M\&P \cite{minier15} have overlooked a fundamental property of the
particle dispersion process. That is in the dynamical system described
by equation (\ref{pem1}), the particle position and velocity are
not independent. This is reflected in the fixed-frame kinetic equation
(\ref{pde1}) through the term $\partial_{x}vp$, which couples the
spatial and velocity distributions of the particles. In the same way,
the distributions of the variables $\widetilde{Z}_{p1},\widetilde{Z}_{p2}$
are coupled in equation (\ref{pde3}). The implication of this coupling
is that fluctuations in particle velocity give rise to fluctuations
in particle position, in addition to the fluctuations in particle
position that arise directly from fluctuations in the fluid force
$\tau_{p}^{-1}\ve{U}_{s}$. In the moving frame it is the fluctuations
in $\widetilde{Z}_{p2}$ (with the +ve eigenvalue, $\omega_{2}$)
via the $+ve$ covariance between between $\widetilde{Z}_{p1}$ and
$\widetilde{Z}_{p2}$, that overcomes the $-ve$ diffusion associated
with $\widetilde{Z}_{p1}$ (in the absence of the coupling). We note,
for instance, that in equation (\ref{pde1}), the particle flux $\ve vp$
integrated over all particle velocities is expressible as a net gradient
diffusion flux ,$\overline{\ve v}p_{r}$ for which the long term ($t\rightarrow\infty)$
particle diffusion coefficient $\varepsilon(\infty)$ in statistically
stationary, homogeneous, isotropic turbulence is given by 
\begin{equation}
\varepsilon(\infty)=\tau_{p}\left\{ \left\langle v^{2}(\infty)\right\rangle +\lambda(\infty)\right\} \label{eq:diffusion coefficient}
\end{equation}
where $\left\langle v^{2}(\infty)\right\rangle $ is the variance
of the particle velocity (which for a Gaussian process is given by
$(\tau_{p}/3)\mathrm{trace}(\bm{\mu}(\infty))$, see e.g equations
(78-79) in \cite{Reeks92}), and $\lambda=(1/3)\mathrm{trace}(\bm{\lambda})$.
This simple relationship clearly identifies the two sources of dispersion
independently, the first from fluctuations in the particle velocity
(the kinetic contribution) and the second term $\mbox{\ensuremath{\lambda}(\ensuremath{\infty)}}$
arising from fluctuations in $\tau_{p}^{-1}\ve{U}_{s}$ (the turbulent
aerodynamic force contribution). We refer to \cite{Reeks91PoF} for
a detailed analysis of how this relationship defines an equation of
state for the particle pressure and where $\left\langle v^{2}(\infty)\right\rangle $and
$\lambda(\infty)$ are more correctly identified as the normal components
of stress tensors. We refer to \cite{Reeks83} on how a proper treatment
of the integrated flux terms in the kinetic equation in inhomogeneous
turbulence gives rise to \emph{turbophoresis}, an important mechanism
for particle deposition (in response the unfounded criticism in both
\cite{minier15,MINIER20161} that the kinetic equation is inappropriate
for modeling particle deposition).

To demonstrate these features in a quantitative way we consider the
simple 2D case examined by M\&P in which $\langle\ve{U}\rangle=\bm{0}$,
and $\widetilde{\ve{Z}}_{p}(0)=\widetilde{\ve{z}}^{0}$ fixed. Then
$\widehat{\ve{\mathrm{a}}}$ is linear in $\widetilde{\ve{z}}$, and
$\overline{\widehat{\mathrm{a}}}_{1}\widetilde{p}_{r}$ involves $\overline{\widetilde{\mathrm{z}}}_{2}\widetilde{p}_{r}=\int{\widetilde{\mathrm{z}}}_{2}\widetilde{p}\,d\widetilde{\mathrm{z}}_{2}$.
This can be expressed in terms of convective and gradient diffusive
fluxes (see \cite{swailes97}) 
\begin{equation}
\overline{\widetilde{\mathrm{z}}}_{2}\widetilde{p}_{r}=\widetilde{\mathit{m}}_{2}\widetilde{p}_{r}-\widetilde{\theta}_{21}\partial_{\widetilde{\mathrm{z}}_{1}}\widetilde{p}_{r}{\color{red}}\label{flux1}
\end{equation}
where $\widetilde{\mathit{m}}_{2}$, $\widetilde{\theta}_{21}$ are
components of $\langle\widetilde{\ve{Z}}_{p}\rangle=\widetilde{\ve{m}}=(\widetilde{\mathit{m}}_{1},\widetilde{\mathit{m}}_{2})$
and $\langle(\widetilde{\ve{Z}}_{p}-\widetilde{\ve{m}})(\widetilde{\ve{Z}}_{p}-\widetilde{\ve{m}})\rangle=\widetilde{\Theta}=(\widetilde{\theta}_{ij})$
satisfying 
\begin{align}
\dot{\widetilde{\ve{m}}} & =\A\cdot\widetilde{\ve{m}}+\widetilde{\ve{k}}\label{m}\\
\dot{\widetilde{\Theta}} & =\A\cdot\widetilde{\mathrm{\Theta}}+\left(\A\cdot\widetilde{\mathrm{\Theta}}\right)^{\mathrm{T}}+\Lambda\label{Theta}
\end{align}
with $\widetilde{\ve{m}}(0)=\widetilde{\ve{z}}^{0}$, $\widetilde{\Theta}(0)=\bm{0}$.
Here $\A=\mathrm{P}^{\mathrm{T}}\cdot\mathbf{A}\cdot\mathrm{P}+\mathrm{R}$,
$\widetilde{\ve{k}}=\mathrm{P}^{\mathrm{T}}\cdot\ve{k}$ with $\ve{k}=(\bm{0},\ve{F}_{\mathrm{ext}})$
and $\mathrm{A}_{11}=\mathrm{A}_{21}=0,\mathrm{A}_{12}=1$, $\mathrm{A}_{22}=-1/\tau_{p}^{\mathrm{St}}$.
equations \eqref{flux1}, \eqref{m}, \eqref{Theta} allow equation
\eqref{pde3} to be written 
\begin{equation}
\partial_{t}\widetilde{p}_{r}=-\partial_{\widetilde{\mathrm{z}}_{1}}\dot{\widetilde{\mathit{m}}}_{1}\widetilde{p}_{r}\ +\ \partial_{\widetilde{\mathrm{z}}_{1}}^{2}\tfrac{1}{2}\hspace{0.2em}\dot{\hspace{-0.2em}\widetilde{{\theta}}}_{11}\widetilde{p}_{r}\label{pde4}
\end{equation}
The net diffusional effect is therefore determined by the particle
diffusion coefficient $\widetilde{D}_{1}(t)$ of the transformed variable
$\tilde{z}_{1}$ (associated with the negative eigenvalue $\omega_{1}$)
and given by 
\begin{equation}
\widetilde{D}_{1}(t)=\tfrac{1}{2}\hspace{0.2em}\dot{\hspace{-0.1em}\widetilde{\theta}}_{11}=(\A\cdot\widetilde{\Theta})_{11}-\frac{1}{2}\vert\omega_{1}\vert\label{eq:D11}
\end{equation}
This shows how the `anti-diffusion' associated with $\omega_{1}$
is offset by the contribution emerging from the flux $\overline{\widehat{\mathrm{a}}}_{1}\widetilde{p}_{r}$
associated with the coupling between $\widetilde{Z}_{p1}$ and $\widetilde{Z}_{p2}$
through their covariance $\widetilde{\theta}_{12}$ in equation (\ref{eq:D11}).

\begin{figure}[h]
\includegraphics[scale=0.5]{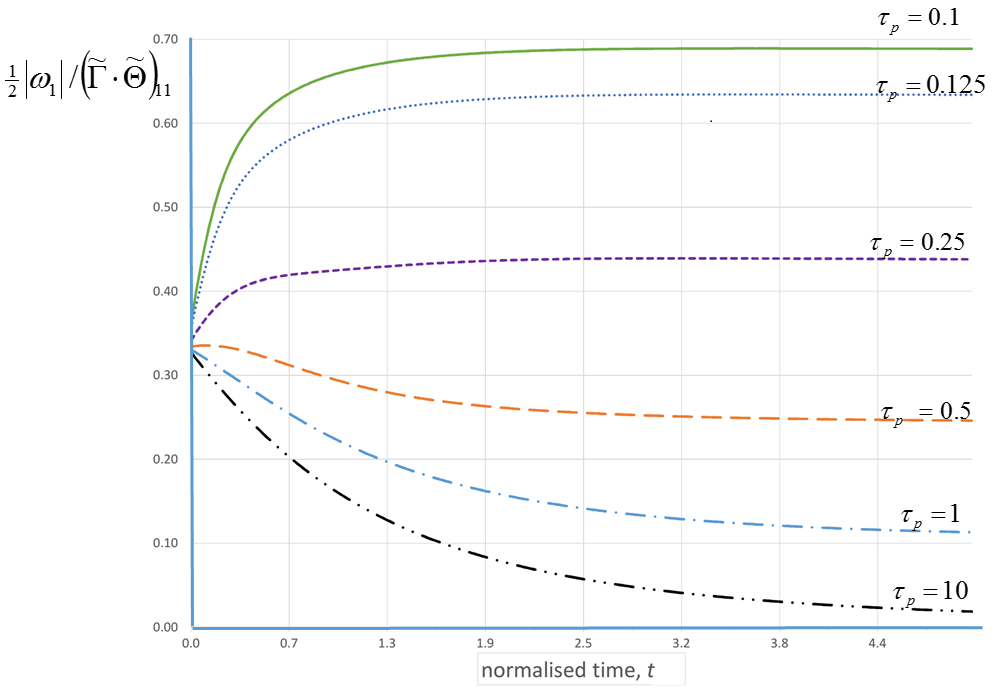} \protect\protect\protect\caption{Plots of $\frac{1}{2}\vert\omega_{1}\vert/(\A\cdot\widetilde{\Theta}{}_{11})$
equation (\ref{eq:D11}) for the ratio of $-ve$ /$+ve$ contributions
to the particle diffusion coefficient $\widetilde{D}_{1}$ of the
transformed variable $\widetilde{Z}_{p1}$(with a $-ve$ eigenvalue)
in the moving frame of reference, as a function of time $t$ for a
range of values of the particle response time $\tau_{p}$. Both $t$
and $\tau_{p}$ are scaled on $T_{L}$, the Lagrangian integral timescale
of the carrier flow measured along a particle trajectory.}
{\label{figure-1}} 
\end{figure}

\begin{figure}
\includegraphics[scale=0.5]{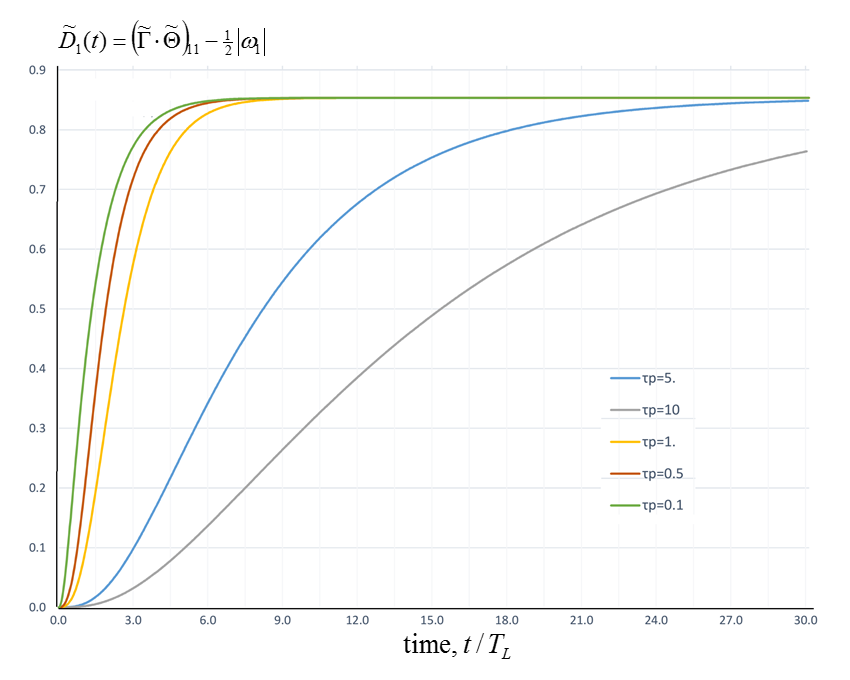} \protect\protect\protect\caption{\label{fig:figure-2}Evolution of the particle diffusion coefficient
$\widetilde{D}_{1}(t)$ evaluated using equation (\ref{eq:D11}) in
the moving frame of reference for a range of values of $\tau_{p}$
(the particle response time normalised on the Lagrangian integral
time scale,$T_{L}$). Time is real time $t$ normalised on $T_{L}.$}
\end{figure}

Figure \ref{figure-1} demonstrates that $0\leqslant\frac{1}{2}\vert\omega_{1}\vert/(\A\cdot\widetilde{\Theta})_{11}\leqslant1$.
The plots, which show the time evolution of this ratio for a range
of values for $\tau_{p}$ (with $\mb{F}_{\mathrm{ext}}=\bm{0}$),
were obtained from closed form solutions of \eqref{Theta}. These
solutions are constructed by noting that $\widetilde{\ve\Theta}=\mathrm{P}^{\mathrm{T}}\cdot\ve\Theta\cdot\mathrm{P}$,
where the covariances $\ve\Theta=\langle{\mb{Z}}_{p}{\mb{Z}}_{p}\rangle$
in the fixed frame are governed by a set of equations analogous to
equations (\eqref{Theta}) \cite{Reeks91PoF} which can be integrated
analytically. We refer to \cite{Reeks91PoF} where analytic solutions
are given for \textbf{$\ve\Theta$} in terms of $\langle\mathrm{U}_{s}^{\prime}(0)\mathrm{U}_{s}^{\prime}(t)\rangle$
the autocorrelation of the carrier flow velocity fluctuations sampled
along particle trajectories. The values of the $-ve$ to $+ve$ ratio
plotted in Figure \ref{figure-1} were obtained using an exponential
decay $\exp\left[-t/T_{L}\right]$ for this autocorrelation. For completeness
we also show in Figure \ref{fig:figure-2} for a similar range of
values of $\tau_{p}$, the evolution of the particle diffusion coefficient
$\widetilde{D}_{1}(t)$ in the moving frame of reference indicating
not only that $\widetilde{D}_{1}\geqslant0$, but also that it reaches
an asymptotic limit that is the same for all $\tau_{p}$. This is
is also true of the particle diffusion coefficient $\varepsilon(\infty)$
in the fixed frame of reference, equation (\ref{eq:diffusion coefficient}).
In particular in the normalised units used to express the values for
$\widetilde{D}_{1}$ in Figure $\ref{fig:figure-2}$, $\varepsilon(\infty)=1$.
This result is universally true for a particle equation of motion
involving the linear drag form in equation (\ref{pem1}) for statistically
stationary homogeneous isotropic turbulence (see \cite{Reeks1977}
where it is $T_{L}$ that depends on $\tau_{p}$). An evaluation of
the asymptotic form of $\langle{\mb{Z}}_{p}{\mb{Z}}_{p}\rangle$ which
is linear in $t$ in this limit shows that 
\begin{equation}
\begin{array}{ccc}
\widetilde{D}_{\textcolor{red}{1}}(\infty)=1/(4-2\sqrt{2})\\
\widetilde{D}_{2}(\infty)=1/(4+2\sqrt{2})
\end{array}\label{eq:D1 asymptotic form}
\end{equation}
and is consistent with the forms for $\widetilde{D}_{1}(t)$ in Figure
\ref{fig:figure-2} obtained by solving a coupled set of equations
(\ref{Theta}) for $\widetilde{\ve\Theta}$. That the asymptotic result
in equation (\ref{eq:D1 asymptotic form}) agrees with the results
in Figure \ref{fig:figure-2} provides not only a check for the analytic
solutions used in Figure \ref{fig:figure-2}, but also a proof that
the $+ve$ contribution to $\widetilde{D}_{1}(t)$ will always outweigh
the $-ve$ contribution in equation (\ref{eq:D11}) (i.e. it applies
to all physically acceptable forms of the autocorrelation for $\mathrm{\ve U}_{s}$,
and not just the decaying exponential form of $\langle\mathrm{U}_{s}^{\prime}(0)\mathrm{U}_{s}^{\prime}(t)\rangle$
that we have chosen to obtain our analytical results).

This must be so for two reasons. Firstly the route involving a solution
of the kinetic equation in the fixed frame of reference and the linear
relationship between the fixed and transformed variables always ensures
a realizable Gaussian distribution for the transformed variables.
Secondly in this calculation this realizability does not itself explicitly
involve or rely in any way on whether one of the eigenvalues $\omega_{i}<0$
and any explicit form for $\langle\mathrm{U}_{s}^{\prime}(0)\mathrm{U}_{s}^{\prime}(t)\rangle$
we might choose, only that the transformation matrix $\ve P$ formed
from the normalised eigenvectors of the diffusion matrix exists and
is well behaved. However the second route via equation (\ref{Theta})
only ensures a realizable Gaussian process if the $+ve$ contribution
to $\widetilde{D}_{1}(t)$ exceeds the $-ve$ contribution. But since
the two methods of calculating $\widetilde{\Theta}$ are in the end
mathematically equivalent to one another, then the $+ve$ contribution
to $\widetilde{D}_{1}(t)$ must always exceed the $-ve$ contribution
in equation (\ref{eq:D11}).

\begin{figure}
\includegraphics[scale=0.5]{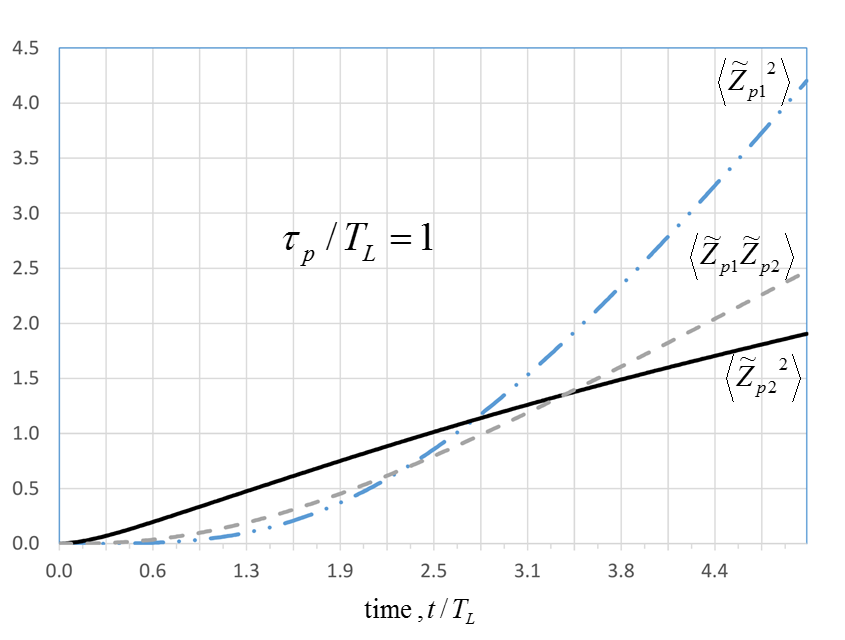}\protect\protect\protect\caption{Moments \label{fig:fig-3}$\left\langle \widetilde{Z}_{pi}\widetilde{Z}_{pj}\right\rangle $
in the moving frame of reference based on the moments $\left\langle \ve Z_{p}\ve Z_{p}\right\rangle $for
$\tau_{p}/T_{L}$ in the fixed frame of reference as solutions of
the fixed frame kinetic equation (\ref{pde1}) or equivalently by
evaluating $\langle{\ve{Z}}_{p}{\ve{Z}}_{p}\rangle$ from solutions
of the particle equation of motion equation (\ref{pem1}).}
\end{figure}

We show the values of the moments $\left\langle \widetilde{Z}_{pi}\widetilde{Z}_{pj}\right\rangle $
in Figure \ref{fig:fig-3} appropriate for the Gaussian function solution
of the kinetic equation in the moving frame (see equation (87) in
\cite{Reeks91PoF}). There is of course no hint of a singularity in
Figure \ref{fig:fig-3}, all 3 moments being smoothly varying, monotonically
increasing in time and linear in time for $t/T\gg1$.

The results also illustrate the now obvious result that, at large
times, the two contributions to the diffusional transport are of the
same order in $t$. The claim in \cite{minier15} that equation \eqref{pde3}
reduces to the form of a backward heat equation because $\overline{\widehat{\mathrm{a}}}_{1}\widetilde{p}_{r}\rightarrow0$
as $t\rightarrow\infty$ is invalid. It fails to acknowledge that
$\omega_{1}\partial_{\widetilde{\mathrm{z}}_{1}}\widetilde{p}_{r}\rightarrow0$
at the same rate. \\
 \\
 Although we have now demonstrated that the transformed kinetic equation
is not ill-posed, we close this section with some comments on M\&P's
use of the Feynman-Kac formula (FKF) and the associated arguments
in \cite{minier15}. In \cite{minier15}, M\&P suggest that equation
(\ref{pde3}) has the structure of a (generalized) Backward Kolmogorov
Equation (BKE), that may be derived from FKF. Noting this, M\&P use
the FKF to construct the solution to equation (\ref{pde3}), using
the terminal condition $\widetilde{p_{r}}(\widetilde{\mathrm{z}}_{1},T)=\Psi(\widetilde{\mathrm{z}}_{1})$,
to obtain ($t\in[0,T]$) 
\begin{align}
\widetilde{p_{r}}(\widetilde{\mathrm{z}}_{1},t)=\Bigg\langle\exp\Bigg[\int_{t}^{T}\partial_{\widetilde{\mathrm{z}}_{1}}\overline{\widehat{\mathrm{a}}}_{1}(\mathcal{X}(s),s)\,ds\Bigg]\Psi(\mathcal{X}(T))\Bigg\rangle_{\mathcal{X}(t)=\widetilde{\mathrm{z}}_{1}},\label{FKf}
\end{align}
where $\mathcal{X}(s)$ is a stochastic process defined through 
\begin{align}
d\mathcal{X}(s)\equiv\overline{\widehat{\mathrm{a}}}_{1}(\mathcal{X}(s),s)ds+\sqrt{\vert\omega_{1}(s)\vert}dW(s),
\end{align}
and $W(s)$ is a Wiener process. M\&P argue that the solution (\ref{FKf})
implies that only ``special'' initial ($t=0$) conditions are permitted
when solving (\ref{pde3}) since (\ref{FKf}) specifies 
\begin{align}
\widetilde{p_{r}}(\widetilde{\mathrm{z}}_{1},0)=\Bigg\langle\exp\Bigg[\int_{0}^{T}\partial_{\widetilde{\mathrm{z}}_{1}}\overline{\widehat{\mathrm{a}}}_{1}(\mathcal{X}(s),s)\,ds\Bigg]\Psi(\mathcal{X}(T))\Bigg\rangle_{\mathcal{X}(0)=\widetilde{\mathrm{z}}_{1}}.\label{FKfic}
\end{align}
From this they conclude that since equation (\ref{FKf}) only applies
for the ``special initial condition'' given by equation (\ref{FKfic}),
then equation (\ref{pde3}) ``is an unstable and ill-posed equation.''
This conclusion is clearly erroneous. Since the FKF employs a terminal
condition in solving the PDE, then provided the PDE is well-posed
as a terminal-value problem, the solution of the PDE at $t=0$ must
of necessity be unique and ``special''. For a well-posed, deterministic
PDE, there exists only one solution at $t=0$ that generates the specified
terminal condition at $t=T$, otherwise solutions to the PDE are not
unique!

If equation (\ref{pde3}) were truly a BKE, then it could indeed be
considered ill-posed since the BKE is in general ill-posed when solved
as a time-forward problem (and equation (\ref{pde3}) is to be solved
as a a time-forward problem with a prescribed initial condition).
However, the important point is that although equation (\ref{pde3})
superficially appears to have the structure of a BKE, it cannot be
considered to be equivalent to a BKE for two reasons. First, as we
have already discussed, the term $\overline{\widehat{\mathrm{a}}}_{1}$
is not a general convection term, but has a specific form since it
is a functional of the solution of the equation (\ref{pde3}). This
is in part a manifestation of the fact that unlike the BKE, equation
(\ref{pde3}) is in fact derived from an underlying process that takes
place in a higher dimensional space (i.e the phase-space). Second,
equation (\ref{pde3}) is associated with a non-Markovian process,
whereas the BKE corresponds to a Markov process. The implication of
this is that equation (\ref{FKf}) cannot, at least formally, cover
the entire solution space of the PDE in equation (\ref{pde3}), since
equation (\ref{pde3}) admits solutions that correspond to non-Markov
trajectories in the space $\widetilde{\mathrm{z}}_{1}$, which equation
(\ref{FKf}) does not account for since it constructs solutions via
a conditional expectation over Markov trajectories. Therefore, in
the general case, the FKF cannot be used to say anything categorical
regarding the solutions to equation (\ref{pde3}).

\vspace{1em}

\section{KINETIC and GLM EQUATIONS}

\label{2}

It has been claimed in recent studies of PDF methods \cite{minier15,MINIER20161},
that the kinetic PDF is the marginal of the GLM PDF. This claim is
based on analysis that purports to show that the dispersion tensors
appearing in a kinetic PDF equation derived from the GLM PDF equation
are `strictly identical' to the corresponding tensors emerging directly
from the kinetic modeling approach. If this is so the claim of ill
posedness of the kinetic equation contradicts the well posedness associated
with the Fokker-Planck equation of the GLM. Of course, as we have
just demonstrated, this claim of ill-posedness is ill founded. Here
we consider the validity of the analysis presented in \cite{minier15}
to demonstrate how the kinetic equation can be derived from the GLM
PDF equation.

The analysis is based on the construction of a closure for $\langle\ve{u}_{s}\mathcal{P}\rangle$
where $\ve{u}_{s}(t;\ve{x})=\ve{U}_{s}(t)-\langle\ve{U}_{s}(t)\vert(\ve{X}_{p}(t)=\ve{x})\rangle$
and $\mathcal{P}(\ve{x},\ve{v},t)=\delta(\ve{X}_{p}(t)-\ve{x})\delta(\ve{U}_{p}(t)-\ve{v}))=\delta(\ve{Z}_{p}(t)-\ve{z})$.
We make the simple observation that the ensemble $\langle\cdot\rangle$
to be considered in this closure involves \textbf{\emph{all}} realizations
of the system being considered. It is not, nor can it be interpreted
as an average over only those realizations in which the trajectories
$\ve{Z}_{p}$ satisfy the end-condition $\ve{Z}_{p}(t)=\ve{z}$. Indeed
this is why $\langle\ve{u}_{s}\mathcal{P}\rangle=\langle\ve{u}_{s}\rangle_{\mathbf{z}}\,p(\ve{z},t)$,
where $\langle\cdot\rangle_{\mathbf{z}}$ denotes an average based
on the sub-ensemble containing only those trajectories satisfying
this end-condition. Although self-evident, this point is missed in
the closure formulated in \cite{minier15}. This closure is constructed
by introducing paths $\bm{\omega}(s)=\bm{\omega}(s;\ve{z},t)$ such
that $(\bm{\omega}(t),\dot{\bm{\omega}}(t))=\ve{z}$. These paths
are used to partition particle trajectories; for a given path $\bm{\omega}(\cdot;\ve{z},t)$
define $\bm{\Omega}_{\bm{\omega}}=\{\ve{Z}_{p}:\ve{X}_{p}(s)=\bm{\omega}(s;\ve{z},t)\}
$. In \cite{minier15} a closure is then considered for the sub-ensemble
$\langle\ve{u}_{s}\mathcal{P}\rangle^{\bm{\Omega}_{\bm{\omega}}}$
over those trajectories in $\bm{\Omega}_{\bm{\omega}}$ (see equation
(39) in \cite{minier15}), and this closure is then integrated over
all paths $\bm{\omega}(\cdot;\ve{z},t)$. Thus, only trajectories
satisfying the specified end-condition $\ve{Z}_{p}(t)=\ve{z}$ have
been taken into account. This is wrong. Moreover, the form of the
closure for $\langle\ve{u}_{s}\mathcal{P}\rangle^{\bm{\Omega}_{\bm{\omega}}}$
is questionable. The Furutsu-Novikov formula is invoked, 
the correct application of this should result in a closure framed
in terms of the two-time correlation tensor $\mathrm{C}(s,s^{\prime};\ve{z},t)=\langle\ve{u}^{\bm{\omega}}(s)\ve{u}^{\bm{\omega}}(s^{\prime})\rangle^{\ve{u}^{\bm{\omega}}}$
of the process $\ve{u}^{\bm{\omega}}(s)=\ve{u}_{s}(\bm{\omega}(s;\ve{z},t),s)$.
However, in \cite{minier15} this is conflated with another correlation,
namely 
\begin{equation}
R(s,\ve{x};s^{\prime},\ve{x}^{\prime})=\langle\ve{u}_{s}(s;\ve{x},t)\ve{u}_{s}(s^{\prime};\ve{x}^{\prime},t^{\prime})\rangle\label{R}
\end{equation}
Again, this is evidently wrong; $\mathrm{C}$ depends on a single
phase-space point, $\ve{z}$, whereas $R$ is defined in terms of
two points $\ve{x}$, $\ve{x}^{\prime}$ in configuration space. Not
only this, the ensembles over which these two correlation tensors
are constructed are different. Finally (and notwithstanding these
apparent oversights), even if the resulting forms of the dispersion
tensors emerging from the construction given in \cite{minier15} were
correct, it is incorrect to claim that these tensors are identical
to those appearing in the PDF equation of the kinetic model. In the
kinetic PDF equation the dispersion tensors are defined in terms of
the basic two-point, two-time correlation tensor of the underlying
fluctuations in the carrier flow velocity field, that is $\mathcal{R}(\ve{x},t;\ve{x}^{\prime},t^{\prime})=\langle\ve{u}^{\prime}(\ve{x},t)\ve{u}^{\prime}(\ve{x}^{\prime},t^{\prime})\rangle$.
This makes no reference to particle trajectories and, therefore, $\mathcal{R}$
cannot be deemed identical to $R$ defined by equation \eqref{R}.

\section{Limitations of the GLM for dispersed particle flows}

\label{3}

\textcolor{black}{That the GLM is a model and not a fundamental theory
of particle dispersion in turbulent flows, is not an issue of critical
concern. Like all models it has its advantages as well as its limitations.
For instance an obvious advantage is that the }GLM PDF\textcolor{black}{{}
includes the flow velocity sampled along a particle trajectory as
an additional statistical variable as well as the particle velocity
and position. So a solution of the }PDF\textcolor{black}{{} equation
in principle contains more information about the dispersion process
than the solution of the kinetic equation. Most noticeably Simonin
and his co workers have used this }PDF\textcolor{black}{{} equation
to formulate transport equations for the density weighted mean flow
velocity $\overline{\ve U_{s}}$ and the particle-flow covariances
and obtained remarkably good agreement with experimental measurement
in numerous particle laden flows including jets and vertical channel
flows \cite{Simonin96b,ReeksSimoninFede}. Van Dijk \& Swailes \cite{vanDijk_Swailes2012}
solved this GLM }PDF\textcolor{black}{{} equation numerically directly
in the case of particle transport and deposition in a turbulent boundary
layer showing the existence of singularities in the near wall particle
concentration. Reeks \cite{Reeks2005} solved this }PDF\textcolor{black}{{}
equation for particle dispersion in a simple shear and obtained valuable
insights into the influence of the shear on the fluid velocity correlations
as well as the dispersion in the streamwise direction which showed
a component of contra-gradient diffusion \cite{Reeks2005}.}\\
 \\
 \textcolor{black}{Our aim here is to point out the limitations of
the GLM for dispersed gas-particle flows that have been ignored in
previous analyses especially in \cite{minier15}, to give} a\textcolor{blue}{{}
}\textcolor{black}{more balanced view of its strengths and weaknesses
when compared to the kinetic approach. We regard these limitations
to be areas for improvement of the model rather than inherent deficiencies.
The advantage of models of this sort is that features inherent in
more fundamental approaches like the kinetic approach can be included
in an }\textcolor{black}{\emph{ad hoc}}\textcolor{black}{{} manner.}\\
 \\
 Central to the formulation of the GLM PDF equations is the need to
model $\dot{\ve{U}}_{s}$ (equations (23), (24), (25) in \cite{minier15}).
By definition 
\begin{equation}
{\color{red}{\normalcolor \dot{\ve{U}}_{s}(t)=\Bigg(\frac{D\ve u_{f}}{Dt}-(\ve{u}_{f}-\ve{U}_{p})\cdot\partial_{\mathbf{x}}\ve{u}_{f}\Bigg)_{\bm{x}=\bm{X}_{p}(t)}}}\label{dotUs}
\end{equation}
with $D\ve{u}_{f}/Dt$ denoting the fluid acceleration field , and
$(\cdot)_{\bm{x}=\bm{X}_{p}(t)}$ denoting that the field variables
inside the parenthesis are evaluated at the particle position. Equation
(\ref{dotUs}) shows that the process $\dot{\ve{U}}_{s}(t)$ is fundamentally
connected to the properties of the underlying flow fields, and as
such is influenced by the spatio-temporal structure of those fields.
This is particularly important since it is known, for example, that
inertial particles interact with the topology of fluid velocity fields
in particular ways, with a preference to accumulate in the strain
dominated regions of the flow \cite{Maxey1987}. Equation (\ref{dotUs})
captures the way in which the process $\dot{\ve{U}}_{s}(t)$ is affected
by the properties of the underlying flow fields. However, in the GLM,
$\dot{\ve{U}}_{s}(t)$ is modeled using a Langevin equation, and as
such, the influence of the spatio-temporal structure of the underlying
fields on $\dot{\ve{U}}_{s}(t)$ is lost. This means then that the
GLM cannot properly capture the role of flow structure on inertial
particle dynamics in turbulent flows, which is known to be very important
for describing the spatial distributions of the particles. In contrast
to this, the kinetic model does capture the role of the spatio-temporal
structure of the flow on the particle motion. For example, the dispersion
tensors $\ve\lambda$, $\ve\mu$ and $\ve\kappa$ capture these effects
through their dependence on the two-point, two-time correlation tensor
of the fluid velocity field. \\

A second, related issue, concerns the handling of the term $(\ve{u}_{f}-\ve{U}_{p})\cdot\partial_{\ve{x}}\ve{u}_{f}$
in the GLM. The role of this term in (\ref{dotUs}) is that it captures
how the particle inertia causes the timescale of ${\ve{U}}_{s}(t)$
to deviate from the Lagrangian timescale of the fluid velocity. For
example, in the limit $\tau_{p}\to0$, one should recover $\dot{\bm{U}}_{s}=(D\ve{u}_{f}/Dt)_{\bm{x}=\bm{X}_{p}(t)}$,
while in the limit $\tau_{p}\to\infty$ (without body forces), one
should recover $\dot{\bm{U}}_{s}=(\partial_{t}\bm{u}_{f})_{\bm{x}=\bm{X}_{p}(t)}$.
In the former case, the timescale of $\bm{U}_{s}$ is the fluid Lagrangian
timescale, whereas in the latter case the timescale of $\bm{U}_{s}$
is the fluid Eulerian timescale. With body forces e.g. gravity , the
timescale of $\bm{U}_{s}$ for inertial particles would also be affected
by the crossing trajectories effect \cite{Wells_Stock83}.

Conventionally, the term $(\ve{u}_{f}-\ve{U}_{p})\cdot\partial_{\ve{x}}\ve{u}_{f}$
is either neglected, such that the Langevin model relates to $\dot{\bm{U}}_{s}=(D\ve{u}_{f}/Dt)_{\bm{x}=\bm{X}_{p}(t)}$,
or else its effect is modeled by making the timescale in the Langevin
model a function of $\tau_{p}$. Both approaches are problematic:
the first because it neglects the effect of inertia on the timescale
which can be strong, the second because one then requires an additional
model for the timescale of ${\bm{U}}_{s}$ as a function of $\tau_{p}$.
In contrast, in the kinetic model, the role of inertia on ${\bm{U}}_{s}$
is formally accounted for, and is an intrinsic part of the model.
In particular, it is captured through the dependence of $\ve\mu$,
$\ve\lambda$ and $\ve\kappa$ on the correlation tensors of the fluid
velocity \emph{field} evaluated along the inertial particle trajectories.

Another implication of the GLM's use of a Langevin equation to describe
${\ve{U}}_{s}(t)$ is that, as is well known, it cannot accurately
describe the Lagrangian properties of the system in the short-time
'ballistic' limit. For example, the second-order Lagrangian structure
function $\langle\|{\ve{U}}_{s}(t+s)-{\ve{U}}_{s}(t)\|^{2}\rangle$
should grow as $s^{2}$ in the limit $s\to0$, whereas a Langevin
equation predicts that it grows linearly in $s$ in the limit $s\to0$.
Interestingly, this very fact has an important bearing on claim of
the exact correspondence of the PDF of the kinetic equations with
the marginal of the GLM PDF. Even aside from other issues, this claim
cannot be correct since the kinetic model would give the correct short-time
behavior for $\langle\|{\ve{U}}_{s}(t+s)-{\ve{U}}_{s}(t)\|^{2}\rangle$
since it allows for the general case where the fluid velocity field
is differentiable in time.

\textcolor{black}{In addition to these points, recent criticism of
the kinetic equation failed to appreciate or show any awareness of
important consistency and invariance principles that were important
guidelines in the construction of the kinetic equation and highly
relevant to the to the limitations and generality of the GLM PDF equations.
The first is that the kinetic equation should generate the correct
equation of state, i.e. the relation between the equilibrium pressure
associated with the correlated turbulent motion of the particles and
their mass density in homogeneous isotropic statistically stationary
turbulence. This can be obtained independently of the kinetic equation
by evaluating the Virial for the particle equation of motion (see
section II in \cite{Reeks91PoF}). This relates the kinematic pressure
$\hat{p}$ to the particle diffusion coefficient $\varepsilon$ via
the particle response time $\tau_{p}$, namely $\hat{p}=\varepsilon\tau_{p}^{-1}$.}

\textcolor{black}{The second important consideration is that the kinetic
equation should satisfy Random Galilean Transformation (RGT) invariance
\cite{Kraichnan65,Reeks92,Frisch95} . In the development of legitimate
closure schemes invariance to RGT is crucial to account for the transport
of small scales of turbulence by the large scales and the $E(k)\sim k^{-5/3}$.
Specifically RGT means applying to each realization of the carrier
flow a translational velocity, constant in space and time but varying
randomly in value from one realization to the next. In Kraichnan\textquoteright s
traditional usage of RGT the distribution of velocities is taken to
be Gaussian for convenience. Clearly the internal dynamics should
be unaffected by this transformation and should be reflected in the
equations that describe the average behavior of the resulting system.
In the case of the case of the kinetic equation the terms that describe
the dispersion due to the aerodynamic driving force and that due to
the transnational velocity should be separate. When the timescale
of $\bm{U}_{s}$ is finite, RGT cannot be satisfied by a PDF equation
with the traditional Fokker-Planck structure. Indeed, RGT invariance
implies that the dispersion tensor $\ve B$ in equation (\ref{pde1})
must have the form given in equation (\ref{B}) \cite{Reeks91PoF},
which is not satisfied in a PDF equation with the Fokker-Planck structure
(for which $\ve\lambda\equiv\bm{0}$).}

\textcolor{black}{This failure to preserve RGT invariance means failure
to reproduce the correct equation of state for the dispersed phase.
In the case of the kinetic equation, it implies that that the dispersion
tensor $\ve B$ in Eq.(\ref{pde1}) must have the form given in Eq.(\ref{B})
for a Gaussian non-white noise process. See \cite{Reeks91PoF} for
the form of the dispersion tensor $\ve B$ for a non Gaussian process
as a cumulant expansion in particle fluid velocity correlations. In
the case of the GLM equations the failure to preserve RGT invariance
is associated with the fluctuating stochastic acceleration field $\dot{\ve U}_{s}^{\prime}=\dot{\ve U}_{s}-\left\langle \dot{\ve U_{s}}\right\rangle $
described by a Fokker Planck equation. It is highlighted in the case
of short term dispersion in e.g homogeneous station ary turbulence
where the GLM predicts an exponential decay for the fluid velocity
autocorrelation (along particle trajectories) and with it a discontinuity
in the slope at $t=0$ and a consequent error in the short term dispersion
of $\mathcal{O}t$ as opposed to $\mathcal{O}t^{2}$. Such a result
cannot arbitrarily be changed since the exponential autocorrelation
is a property of the white noise based GLM equation for all time. }

\textcolor{black}{This has some bearing on the equivalence of the
two approaches, since the kinetic approach does not have this limitation
and correctly predicts the short term diffusion. So whereas in the
GLM, the form of the particle-flow correlations are calculated and
an intrinsic part of the model, in the kinetic equation these are
prescribed or calculated using independent knowledge of the statistics
of the carrier flow field and a relationship between Eulerian and
Lagrangian correlations. As pointed out in \cite{Reeks2005} in the
case of dispersion in a simple shear flow, if the statistics of the
fluid velocity along a particle trajectory are assumed derivable from
a Gaussian process and the fluid velocity correlations as a function
of time are taken to be the same in either case, then the two approaches
are identical, but only then. Whilst in the kinetic equation one is
free in principle to choose whatever is physically acceptable for
the fluid particle correlation, the problem remains one of calculating
carrier flow velocity correlations along particle trajectories, given
the underlying Eulerian statistics of the carrier flow velocity field. }

\subsection*{The kinetic equation for non-linear drag}

In closing this section, we wish to address the numerous claims made
that the kinetic approach is limited in its application to situations
where the drag force is linear in the relative velocity between particle
and fluid. This is not correct. We refer the reader to section III
in \cite{Reeks92} on the particle motion which specifically deals
with the treatment of non linear drag and how it is used to evaluate
the convective and dispersive terms in the kinetic equation. In particular
the mean and fluctuating aerodynamic driving force are expressed in
terms of the particle mean density weighted particle velocity $\overline{\ve v}(\mathbf{x},t)$
and incorporated into the particle momentum equations by suitably
integrating the kinetic equation over all particle velocities. We
refer also to \cite{Reeks1980} where using the kinetic equation for
nonlinear drag, an evaluation is made of the long term diffusion coefficient
for high inertial particles in homogeneous isotropic statistically
stationary turbulence.

\section{Summary and Conclusions}

This paper is about well-posedness and realizability of the kinetic
equation and its relationship to the GLM equation for modeling the
transport of small particles in turbulent gas-flows. Previous analyses
\cite{minier15,MINIER20161} claim that the kinetic equation is ill-posed
and therefore \emph{invalid} as a PDF description of dispersed two-phase
flows. Specifically, it is asserted that the kinetic equation as given
in equation \eqref{pde1} has the properties of a backward heat equation
and as a consequence its solutions will in the course of time exhibit
finite-time singularities. The justification for this claim is based
on an analysis centered around the observation that the phase space
diffusion tensor $\ve{B}$ in equation (\ref{pde1}) is not positive-definite
but possesses both negative and positive eigenvalues. So we have examined
the validity of assumptions that lead to this conclusion and in particular
the form of the kinetic equation in a moving frame where the PDF $\widetilde{p}(\widetilde{z}_{1},\widetilde{z}_{2},t)$
refers to that of transformed variables $\widetilde{z}_{1},\widetilde{z}_{2}$
measured at time $t$ along the principal axes of $\ve B$ (see equation
 (\ref{eq:transform})). Based on the negative diffusion coefficient
in the transformed PDF equation \eqref{pde3} for the marginal distribution
$\widetilde{p_{r}}(\widetilde{z}_{1},t)$, these studies seek to show
that this equation (and so also equation (\eqref{pde1})) is ill-posed.
However a fundamental error is made by assuming incorrectly that the
term $\overline{\widehat{\mathrm{a}}}_{1}$ in equation (\ref{pde3})
is wholly convective when in fact it is a density weighted variable
which because $\widetilde{z}_{1}$ and $\widetilde{z}_{2}$ for the
particle motion are coupled in phase space, means that $\overline{\widehat{\mathrm{a}}}_{1}$
has a gradient diffusive component with a $+ve$ diffusion coefficient
which offsets the component in equation (\ref{pde3}) with a $-ve$
diffusion coefficient. More particularly, we showed that the solution
is a Gaussian distribution with covariances that are the solutions
of a set of coupled equations in equations (\ref{m}), (\ref{Theta}).
Based on these solutions, the resultant convection-gradient diffusion
equation for $\widetilde{p}_{r}(\widetilde{z}_{1},t)$ is given by
equation (\ref{pde4}) with a diffusion coefficient $\widetilde{D}_{1}(t)$
given by the sum of the $+ve$ and $-ve$ contributions defined in
equations (\ref{eq:D11}). Using an exponential decaying autocorrelation
of the fluid velocity measured along a particle trajectory, we obtained
analytic solutions for the $+ve$ and $-ve$ components of $\widetilde{D}_{1}$
which show that the $+ve$ component \textbf{\emph{always}} outweighs
the $-ve$ component and that $\widetilde{D}_{1}$ is crucially always
$+ve$. The corresponding $+ve$ values of $\widetilde{D}_{1}$ are
shown in Figure \ref{fig:figure-2} which indicate that $\widetilde{D}_{1}(t)$
approaches an asymptotic value that is independent of the particle
response $\tau_{p}$, evident from asymptotic expression given in
equation (\ref{eq:D1 asymptotic form}). Significantly we were able
to show that this was a general result for all realizable forms for
the flow velocity autocorrelation along particle trajectories and
as a consequence the kinetic equation is not ill-posed. 

Finally, in the course of our examination of the analysis of ill-posedness,
we pointed out a number of issues with the use of the Feynman-Kac
formula (FKF). The application of the FKF to equation (\ref{pde3})
is problematic because equation (\ref{pde3}) is not really a Backward
Kolmogorov Equation. Furthermore, the claim that the FKF solution
to equation (\ref{pde3}) implies that the kinetic equation is only
solvable for special initial conditions is erroneous. The FKF employs
a terminal condition, and therefore there can be only one possible
``initial condition'', or else solutions to equation (\ref{pde3})
would not be unique. 

Another important issue was the claim made in \cite{minier15} that
the kinetic equation can be derived from the GLM PDF equation. That
in fact the GLM is a more general approach than the kinetic approach.
We showed that this is not the case, that the assumptions made regarding
the averaging process lead again to a fundamental error in the closure
approximation that negates this claim.

In the final part of our analysis we sought to give a more balanced
appraisal of the benefits of both PDF approaches\textcolor{black}{{}
and in particular to point out the limitations of the GLM for gas-particle
flows that have previously been ignored. We regarded these limitations
to be areas for improvement of the GLM rather than inherent deficiencies.
As we pointed out, the value of models of this sort is that features
inherent in more fundamental approaches like the kinetic approach
can be included in an ad-hoc manner}. None the less there were terms
that were fundamental to the modeling like the fluctuating convective
strain rate contribution which had been ignored but which contained
valuable information on the relationships between Lagrangian and Eulerian
timescales and the dependence on particle inertia. We suggested how
additional features like particle clustering and drift in inhomogeneous
turbulent flows particularly in turbulent boundary layers might be
included in the model to make it more complete. This is one of the
ways that the kinetic approach can support the PDF dynamic model by
giving specific formulae for these additional features.

\bibliographystyle{plain}
\nocite{*}
\bibliography{biblio_PRE_revised}

\end{document}